# Design and Testing of Dimes Carbon Ablation Rods in the DIII-D Tokamak

Dmitri M. Orlov[1,*], Michael O. Hanson[2], Jason Escalera[2], Hadith Taheri[2], Caitlin N. Villareal[2], Daniel M. Zubovic[2], Igor Bykov[3], Evdokiya G. Kostadinova[4,5], Dmitri L. Rudakov[1], and Maziar Ghazinejad[2]

[1]Center for Energy Research, University of California San Diego, La Jolla, CA 92093-0417
[2]Department of Mechanical and Aerospace Engineering, University of California San Diego, La Jolla, CA 92093-0411
[3]General Atomics, San Diego, CA 92186-5608
[4]Center for Astrophysics, Space Physics & Engineering Research (CASPER), Baylor University, Waco, TX 76798
[5]Department of Physics, Auburn University, Auburn, AL 36849

*email: orlov@fusion.gat.com

**ABSTRACT**

We present the design of ATJ graphite rods developed for ablation experiments under high heat flux (up to 50 MW/m$^2$) in the lower divertor of the DIII-D tokamak [1], a magnetic plasma confinement device. This work is motivated by the need to test ablation models relevant to carbon-based thermal shields used in high-speed spacecraft atmospheric entries, where the heat fluxes encountered can be comparable to those achieved in the DIII-D divertor plasma. Several different designs for the flow-facing side of the rod are analyzed, including "sharp nose," "blunt," and "concave". The last shape is studied for its potential to lower heat fluxes at the rod surface by increased radiation from trapped neutrals and reduced parallel plasma pressure. We also analyze the possibility of applying a thin (approximately 30 microns) layer of silicon carbide (SiC) to the exposed part of several carbon ablation rods to benchmark its erosion calculations and lifetime predictions. Such calculations are of interest as SiC represents a promising material for both thermal protection systems (TPS) and a fusion plasma-facing material (PFM). Preliminary results from the DIII-D rod ablation experiments are also discussed.

## 1. INTRODUCTION

Space exploration involves the quest for knowledge of life: its origins, evolution, and spread throughout the Universe, as well as its future beyond Earth. A crucial factor enabling the Earth's habitability was the transport of volatiles towards the inner Solar System, which likely resulted from the formation of Jupiter [2]. The 2019 National Academy of Sciences (NAS) report, *An Astrobiology Strategy for the Search for Life in the Universe* [3], recognized that understanding such processes in the Solar System's early evolution and exploring the habitability of nearby bodies, including Mars and the icy moons of gas giants, are key science drivers for future missions. In addition to similar objectives, NASA's 2018 Strategic Plan [4] identified as a major goal extending human presence deeper into space and to the Moon for sustainable long-term exploration and utilization. The achievement of these ambitious goals in space exploration requires challenging missions, which may involve high-speed hyperbolic re-entries into Earth's atmosphere and entry missions to gas giants. This motivates the desire to test and adapt thermal protection systems (TPS) for high-enthalpy entries and optimize them for mass savings, e.g., for the sake of higher scientific payload masses. Although modeling and development of TPS technology are ranked as a high priority by NASA [5], utilizing ground test facilities has proven difficult and expensive due to the high-heat environments associated with high-enthalpy missions; conditions hard to reproduce on ground.

Future space exploration goals will require high-speed hyperbolic entries needed for lunar returns as well as missions to Venus, Titan, and the gas giants, which become increasingly challenging in high-gravity and thick atmosphere conditions. Present-day heat shield materials are not suitable for the variety of scenarios required by such missions. For example, the heat load predicted for equatorial entry in Jupiter's atmosphere is ~1 GJ/kg, which yields a TPS mass fraction of 50-100% for any examined material up to date. An even higher percentage is required for off-equatorial entries, invalidating the use of these ablators. Therefore, heat shield ablators with improved properties are sorely needed. The heat shielding materials' main functions for high enthalpy entries are heat dissipation and thermal protection of the vehicle's inner parts, while adding as little as possible to the vehicle's mass. However, testing and modeling of material performance in this regime have historically been challenging due to the lack of adequate ground testing facilities.

Here we report on the design of ATJ graphite rods developed for ablation experiments under high heat flux in the lower divertor of the DIII-D tokamak. As DIII-D low-confinement (L-mode) discharges yield stable plasma and high heat fluxes (up to 50 MW/m$^2$), such experiments present a unique opportunity for examining plasma-materials interactions in space-relevant conditions. DIII-D provides sufficiently long plasma discharges (~3-4 s of flat top conditions) with well-controlled, stable L-mode edge plasma conditions. The resulting heat flux and plasma flow speeds (~10$^4$ m/s) are similar to those experienced during atmospheric entries. In addition, DIII-D has one of the strongest research programs in the area of Plasma-Material Interactions (PMI), owing to the versatile suite of diagnostic tools, including the Divertor Materials Evaluation System (DiMES) [6, 7], a removable sample manipulator. The ablation experiments discussed here focus on exposures of carbon rods to high heat flux conditions using DiMES. The designed system allows for post-mortem analysis of mass recession rates and surface roughness, the results of which will be presented at this conference.

This project aims to lay the foundations of a comprehensive study on ablative materials by developing engineering equations and testing them in the extreme conditions of the DIII-D tokamak. The goal is to determine the fundamental processes expected to dominate material ablation for the extreme conditions of high-enthalpy entry missions, as well as initiate the performance evaluation of different ablative materials. This, in turn, will assist in forming the requirements needed for future ground testing facilities.

## 2. MATERIALS AND METHODS

### 2.1 Materials selection

Material selection for the rod samples determines both the constraints of the experimental design and the appropriate form of ablation equations used for the modeling. Key criteria for the material include even ablation, sufficient strength to avoid breaking, and safety for use in the DIII-D tokamak plasma. Carbon-based materials have historically been used as both tokamak plasma-facing materials and thermal protection systems due to their low weight and resistance to thermal shock. One type of carbon-based material is ATJ graphite, which is a very fine-grained graphite with uniform, isotropic thermal properties. This allows for a more even ablation rate, as the material does not have pockets of varying density. ATJ grade graphite has a high strength and small grains that can be machined to very tight tolerances and sharp details with a fine surface finish [8]. These structural features also reduce the risk of rods breaking off during the plasma discharge, which will introduce an undesired source of carbon impurities at an unpredicted location in the vacuum vessel. ATJ graphite also has a unique set of thermal properties, including low thermal expansion and high thermal conductivity, which allows for favorable resistance to thermal shock. This makes ATJ graphite a robust choice of material that is resistant to the high heat fluxes anticipated during the experiments. ATJ graphite has a well-documented past in DIII-D and is used as the material for most of the plasma-facing wall tiles inside the vacuum chamber. Due to its frequent use in the DIII-D tokamak, there are established upper limits for how many carbon impurities can be introduced in the plasma before disrupting the system. These requirements pose constraints on the rod sample design and exposure times during the experiments.

Another material worth exploring is silicon carbide (SiC), which is of interest to both the aerospace community and the nuclear fusion community. SiC is useful in aerospace applications due to its anti-oxidizing properties [9]. Carbon oxidizes at high temperatures, which reduces its strength and could decrease the performance of spacecraft heat shields. Adding a layer of SiC over a core of carbon-based material, such as graphite, helps protect the substrate from oxidation and assists retention of its material properties. Silicon carbide has well-documented and uniform material properties. In addition, coating ATJ carbon rods with a thin layer of SiC (~30 μm thick) ensures that the maximum amount of Si deposited into the DIII-D plasma during experiments represents a low risk of disrupting the discharge. Thus, in the following discussion we focus on ATJ graphite and SiC as the leading materials to be used in this experimental study of ablation in high heat flux environments [10,11].

### 2.2 Selection of Geometries

#### 2.2.1 Conceptual design

The ablation rods discussed here can be integrated into a DiMES head and inserted in the lower divertor region of DIII-D. The typical DiMES head can hold up to seven samples. However, for the ablation rod experiment, only three rods are inserted at a time, as shown in Figure 1, in order to eliminate potential shadowing effects. These rods vary in geometry and are placed at the same toroidal angle in DIII-D, but at different radial positions. In this study, the ablation of three different geometries is considered: blunt, sharp-nosed, and concave. As the sharp nose and concave rods are orientation-dependent, the rods are snugly fitted in the holder to fix their orientation with respect to the plasma flow and prevent rotational motion. Implementation of a more elaborate system to control the orientation of the rods will be considered for future experiments.

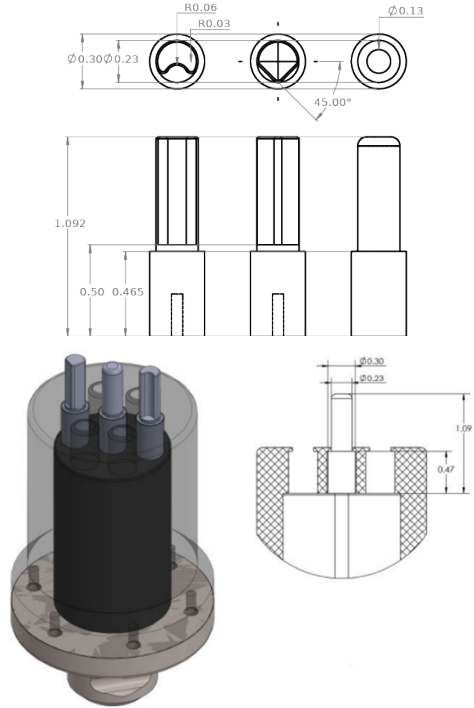

**Figure 1**: (top) An engineering drawing of the blunt, sharp-nose and concave rods showing vertical and horizontal cross-sections. (bottom) A rendered model of the DIII-D DiMES head system with a set of three carbon ablation rods inserted along the line perpendicular to the plasma flow to avoid shadowing. Units are in inches.

### 2.2.2 Shapes

Rods of varying shapes (blunt, sharp-nose, and concave) were designed to observe the impact of geometry on the mass loss rate. Each of the considered rod geometries has a different projected area facing the incoming plasma flow.

The blunt geometries are simple cylindrical rods with rounded ends (to reduce heat concentration at sharp edges), square pillar type, and a semi-cylindrical shape for the flow facing surface. A variation of the cylindrical rod was designed with a smaller exposed diameter to allow for the application of a SiC coating. The blunt geometry rod is frequently used in ablation experiments, which allows for comparison to other ablation studies. The symmetric shape of this rod also makes it the most robust design.

The high velocities found in the divertor region prompted the idea of integrating a sharp-nosed, aerodynamic design to evaluate the ablation rate in comparison to the blunt geometries. The considered designs for the sharp-nose geometry were an airfoil profile rod and a triangular rod, each with varying angles and volumes to allow for control over the surface area exposed to flow. The final sharp-nosed design selected for the DIII-D experiments is a triangular rod with a 45-degree angle of incidence to the incoming flow, which mimics hypersonic vehicle profiles.

The third geometric design is a concave-shaped rod, which allows for exploring the possibility of plasma detachment due to neutrals trapped in the cavity. Such phenomena have been studied in DIII-D using the Small Angle Slot divertor [12] and have potential impact on the ablation rate. With a concave geometric feature facing the flow within the tokamak, we would like to investigate if the cavity "traps" a pocket of less energetic neutrals that serve to protect the main body from high heat flux. Similar to the other geometric rods, the concave rods can vary in surface area facing the flow and volumetric size.

### 2.3 Thermal Analysis

Thermal analysis is needed to estimate the expected heat distribution on the rod due to the applied heat flux. Using the simulation program Ansys 2020 R2, the time at which the rod will reach ablation temperatures can be estimated, which also allows for estimation of the total mass loss rate during the exposure. The thermal expansion of the rod should also be taken into consideration because, as the temperature of the rod increases, the expanding material can potentially crack the DiMES casing.

The ablation rods will experience conduction, convection, and radiation, but due to the high heat flux and temperatures reached, conduction and radiation are the main sources of heat transfer. To estimate the heat flux incident on the rod surface, a UEDGE simulation [13] was performed for a DIII-D reference L-mode discharge 170837. The output of this simulation is a 2D axisymmetric plasma fluid solution for the densities, temperatures, flow velocities in the DIII-D plasma edge, including the scrape-off layer and the private flux region in the lower divertor.

To start the thermal analysis on Ansys, the temperature-dependent material properties of ATJ graphite were tabulated on the Ansys Workbench 2020 R2. In this analysis, we assume that (1) the heat is transferred to the rod through conduction and only to the flow-facing side, as the flux experienced by the other side is an order of magnitude smaller, (2) there are no losses from the rod other than conduction through the DiMES assembly and thermal radiation, (3) the bottom of the DiMES casing is exposed to ambient air which acts as a heat sink, and (4) the rod radiates to the vessel walls, which are at ambient temperature. We neglect the losses via convection due to the vacuum environment. The heat flux is varied over the upstream surface of the rod using two methods. First, the incoming heat flux is modeled by a $\cos(\theta)$ distribution, where $\theta$ is the angle between the incoming flux and a vector normal to the rod surface. This is to ensure that the incoming parallel heat flux is accurately mapped to the curved surfaces of the rods. Next, horizontal cuts are made such that the top ¼ of the rod receives the maximum heat loading, the bottom ¼ receives the minimum loading, and the middle portion of the rod receives the averaged loading. Results from this analysis are summarized in section 3.2.

### 2.4 Ablation Modeling

We calculate the mass loss rate of the carbon sample as the ratio of heat flux incident to the surface of the sample to carbon heat of ablation

$$\Delta m_s = \frac{(-q_r - q_c)}{\Delta H_a} \approx \frac{q_\parallel}{\Delta H_a}. \quad (1)$$

During a typical atmospheric entry, the total heat flux toward the surface of the sample is a combination of radiative heat flux $q_r$ and convective heat flux $q_c$. An important assumption in the present calculation is that the total heat flux to the material surface in the ablation experiment is equal to the heat flux parallel to the magnetic field in DIII-D, i.e., $-q_{rw} - q_{cw} \approx q_\parallel$, where

$$q_\parallel = \gamma c_s n_e k_B T_e. \quad (2)$$

Here $\gamma \approx 7$ is a sheath transmission factor, empirically obtained for DIII-D in [14], the speed of sound is $c_s = sqrt\left[\frac{k_B(T_e+T_i)}{m_i}\right]$, $k_B$ is the Boltzmann constant, and $n_e$ and $T_e$ are the electron density and temperature, respectively, at the edge of the plasma sheath. The ion temperature $T_i$ is assumed approximately equal to the electron temperature. Here we consider two models, one by Park [15] and another by Matsuyama et al. [16], which were previously used to calculate carbon ablation during the Galileo probe entry into Jupiter's atmosphere. Thompson scattering measurements of electron temperature, electron density, and incident heat flux to the sample surface during DIII-D discharge 170837 are used as inputs for both models. In the Park model [15], the carbon heat of ablation in [$J\ kg^{-1}$] is given by

$$\Delta H_a = [23307 - 0.825(T_w - 4000)], \quad (3)$$

where $T_w$ is the temperature at the sample wall, which we assume to be equal to the sublimation temperature of carbon $T_w[K] = T_s[K] = 3800\ K$.

In the Matsuyama model [16], the carbon heat of ablation in [$MJ\ kg^{-1}$] is given by

$$\Delta H_a = 28 - 1.375\ log\ p_s + 27.2(log\ p_s)^2, \quad (4)$$

where the pressure $p_s$ on the surface of the material is assumed equal to the vapor pressure of carbon $p_s[atm] = 9.86923 \times 10^{-6} p_v[Pa]$. The vapor pressure can be calculated from the sublimation temperature, using the following equation $p_v = 10^{14.8 - \frac{40181}{T_s}}$ (as discussed in [17]). Results from the ablation analysis are summarized in Section 3.3.

### 2.5 Thermal Expansion

One of the most undesirable complications during the ablation experiment would be if the casing holding the rods were to crack. The most likely cause for such cracking would be via the thermal expansion of the rod's base causing stresses on the DiMES casing. To account for this possibility, equation (5) was used to estimate the increase in the diameter of the rod's base due to the increase in temperature during exposure.

$$\Delta D = \alpha D_0 (T_f - T_i) \quad (5)$$

Here, $\Delta D$ is the change in diameter, $\alpha$ is the coefficient of linear expansion, $D_0$ is the diameter at the initial temperature, and $T_f/T_i$ are the final and initial temperatures, respectively. Prior to the DIII-D experiments, the base of the rod was reduced in diameter to account for this expansion, mitigating the potential for this failure mode.

### 2.6 Current-Induced Force Analysis

A force analysis allows us to determine the magnitude of the current-induced stresses on the rod after the onset of ablation. The magnetic field was decomposed into its toroidal ($B_\phi$) and poloidal ($B_\theta$) components, and the cross (vector) product was calculated with the current flowing through the rod using equations (6) and (7). Given the location and vertical orientation of the rods on the shelf of DIII-D, the vector product of the rod current J and the poloidal component of the B-field can be written as JxB$_R$ where B$_R$ is the radial component of the poloidal B-field.

$$F_R = (JxB_\phi)L_{Rod} \quad (6)$$
$$F_\phi = (JxB_R)L_{Rod} \quad (7)$$

In the tokamak, the plasma flows throughout the inner chamber during a discharge. As a simplification, we assume that the ions and electrons hitting the walls of the chamber are recycled - they recombine and return to the plasma. However, the electrons are moving at a much faster speed than the ions, leading these electrons to build up at the surface of the wall and create a negative charge. This imbalance causes a Debye sheath which acts as a barrier along the surface of the wall that repels electrons due to the build-up of the negative charge. This decrease in electron flux induces a current. The current that flows normally through the rod is on the order of the ion saturation current. However, after the onset of thermionic emission, the Debye sheath locally collapses, and conservative estimates place the current on the order of the electron saturation current. Due to the magnetic field in the tokamak, a JxB force is induced and needs to be accounted for in our stress analysis of the carbon rods.

SolidWorks simulations were used to predict the maximum current the rod can withstand before breaking. The simulation was performed on the two-centimeter-tall angular rod, the weakest of the three rod geometries, in order to predict the worst-case scenario for the current (as discussed in Section 3.5.)

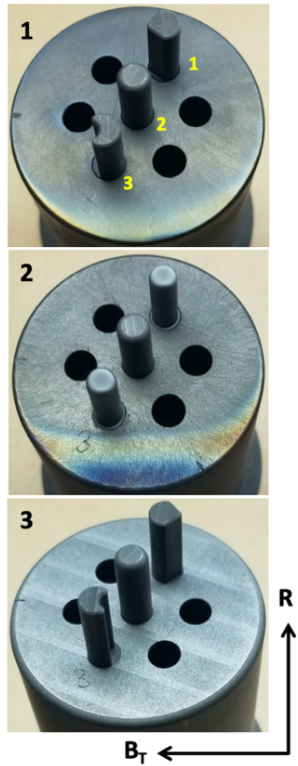

**Figure 2.** Assembly of the three DiMES heads for the experiment 2019-08-04 performed on DIII-D. The layout corresponds to Table 1. The direction of toroidal magnetic field "$B_T$" in DIII-D and radial direction "R" are shown for reference.

## 3. RESULTS AND DISCUSSION

### 3.1 Selection of Geometries

After running thermal and stress analysis from the provided discharge conditions on the rods, it became evident that volumetric size and lengths of the exposed portions of the rod protruding from the DiMES port would experience high heat flux and drive large currents through them. As such, the geometric designs with small volumes had been eliminated along with any sharp edges that experience high heat concentration and static compression/tension forces from the current. These considerations led us to reduce the exposed rod length from 2 cm (initially proposed) down to 1.5 cm, which substantially decreased the expected current-induced forces. However, some of the analysis presented below is focused on the initial 2 cm design.

| DiMES head # | Rod position #1 | Rod position #2 | Rod position #3 |
|---|---|---|---|
| 1 | Backward wedge 1.1<br><br>Pre (g) = 1.7479 | Round 1.2<br><br>Pre (g) = 1.8370 | Backward concave 1.3<br><br>Pre (g) = 1.6496 |
| 2 | Round + SiC 2.1<br><br>Pre (g) = 1.6541 | Round 2.2<br><br>Pre (g) = 1.8364 | Round + SiC 2.3<br><br>Pre (g) = 1.6491 |
| 3 | Wedge 3.1<br><br>Pre (g) = 1.7469 | Round 3.2<br><br>Pre (g) = 1.8357 | Concave 3.3<br><br>Pre (g) = 1.6477 |

**Table 1.** Layout of the three DiMES head assemblies with ablation rods in positions 1,2 and 3. The mass of the rod samples before the experiment (Pre) is given in grams within 0.1mg accuracy.

### 3.2 Thermal Analysis

Using UEDGE simulation results and the location of the DiMES port in DIII-D, the heat flux incident to the surface of the rod was estimated. The rods, which will extend approximately 2 cm into the plasma, will experience a wide range of heat fluxes over their length, ranging from around 27.8 MW/m$^2$ at the top to around 7.79 MW/m$^2$ at the bottom.

Based on the results from a transient thermal analysis on Ansys, a 2 cm tall rod tip would experience a temperature of approximately 4100°C at the end of a 4-second exposure. The temperature distributions are shown in Figure 3. The time series of the temperature reveals that the rod reaches the carbon sublimation temperature of 3800K at approximately 1.2 seconds. After this point, it is assumed that the ablation cloud shields the rod from further incoming plasma flux and therefore maintains this temperature while ablating for the remaining 2.8 seconds of the total 4 second exposure.

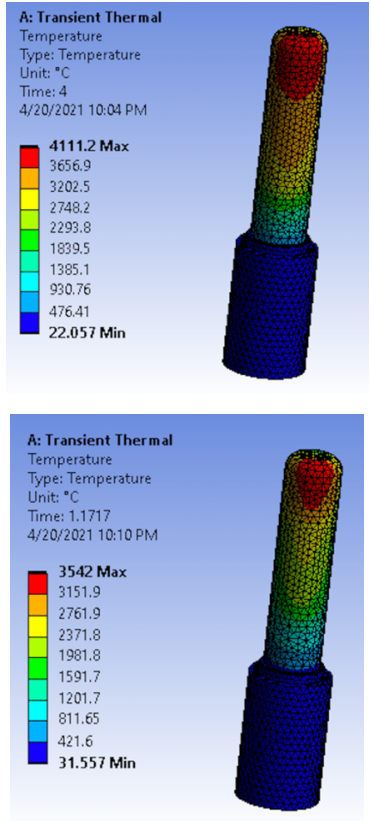

**Figure 3.** (top) Temperature distribution of the rod at the end of a 4 second exposure. (bottom) Temperature distribution of the rod when it reaches ablation temperature.

### 3.3 Ablation Modeling

While the mass loss rate predicted from the Park model (Equation (3)) is about twice as high as the mass loss rate predicted by the Matsuyama model (Equation (4)), both models yield $\Delta m_s < 1\ gs^{-1}cm^{-2}$. Figure 4a shows the expected mass loss rate, calculated using the Park model, while figure 4b shows the expected mass loss rate using the Matsuyama model. Both calculations use $q_\parallel$ measured during the DIII-D discharge 170837. Specifically, assuming a flat flow-facing surface, the Matsuyama model results in a heat of ablation of $\Delta H_a = 45.6\frac{MJ}{kg}$ and a mass loss rate of $\Delta m_s = 0.06\frac{g}{cm^2 s}$ for the plasma conditions at the location of the rods. To obtain the mass loss rate in terms of rod geometry, the red region in Figure 3a is assumed to be the area undergoing ablation. This area approximately represents the top quarter of the exposed portion of the rod and a ~90° arc. Using these values, the ablated area is 0.419 $cm^2$. Multiplying this with $\Delta m_s$ from above, the expected mass-loss rate is 26 mg/s. As the thermal analysis determined that the rod would ablate for ~2.8 seconds, it is expected that 73 mg or ~8% of the total rod will ablate over the exposure time. This suggests that the carbon rod samples in our experiment will only partially ablate if exposed to discharge conditions similar to shot 170837 for exposure times of 3-4 s, as shown in Figure 4.

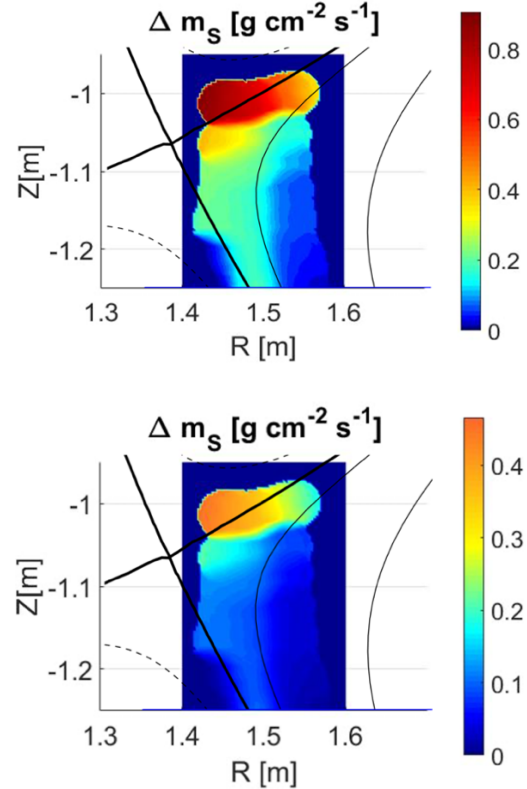

**Figure 4.** Mass loss recession rate predicted using Thomson scattering data from DIII-D discharge 170837 with (top) the Park model and (bottom) the Matsuyama model. Locations of the plasma separatrix (bold black lines) and several constant magnetic flux surfaces in the scrape-off layer and private flux regions are overlaid for reference. The DiMES port with the rods is located at R = 1.46-1.51 m, Z = -1.25 m.

### 3.4 Thermal expansion

Based on the thermal analysis, the base of the rod that is in contact with the DiMES port reaches a temperature of 500 °C at the end of the 4 second exposure time. To ensure that rod expansion will not cause cracking of the DiMES port and to allow for safety factor, it was assumed that the temperature would reach 1000 °C. Using Equation (5), the thermal expansion of the rod's base is calculated. In this equation, the initial diameter of the rod's base is $D_0 =$ 6mm, the initial temperature of rod's base is $T_i = 25$ °C, the final temperature of the rod's base is $T_f = 1000$ °C, and coefficient of thermal expansion at final temperature is $\alpha = 3.7 \times 10^{-6}$ °C$^{-1}$.

Based on these assumptions, it is expected that the rod's base will expand approximately 0.022 mm due to the exposure. To increase the safety factor and ensure that the DiMES port casing will not crack, the rod was undersized by 0.15 mm. As a result, a pin was designed to secure the orientation of the rod and prevent rotation due to the incoming plasma flow.

### 3.5 Current-Induced Force Analysis

Based on the current-induced force modeling, the angular rod was considered the least robust geometry due to its smaller volume and more vulnerable, sharper edges. Table 2 shows the upper limits of the uniform current allowed before failure for each geometry (Round, Concave, Wedge) and three different exposed lengths.

| Design | Maximum uniform current before failure |
|---|---|
| Round (2cm) | 450 A |
| Round (1.5cm) | 555 A |
| Round (1cm) | 645 A |
| Concave (2cm) | 290 A |
| Concave (1.5cm) | 320 A |
| Concave (1cm) | 370 A |
| Wedge (2cm) | 280 A |
| Wedge (1.5cm) | 290 A |
| Wedge (1cm) | 350 A |

**Table 2:** Maximum uniform current before failure tabulated for each of the three geometries for each considered length. This current translates to a toroidal and radial force via Equations 6 & 7.

The bottom of the rod was specified as fixed geometry in our model since the actual rod will be secured inside the DiMES port at this location. While the toroidal and poloidal magnetic fields are assumed fixed to values typical of L-mode discharge, the current was varied until the resulting stresses became greater than the material strength. It was assumed that the current would build linearly from the top to the bottom of the rod. The worst-case scenario of such nonuniform current that did not produce stresses greater than the 3.7 ksi tensile strength of ATJ graphite was found to be 565 Amps. In our analysis, it was determined that the other two rod designs would be able to withstand this current worst-case scenario without failing. However, the forces produced by the 565 Amp current yielded a factor of safety not much higher than one.

Based on the thermal and stress analysis on the rods, it was determined that volumetric size and lengths of the exposed portions of the rod protruding from the DiMES port would experience high heat flux and drive unacceptably large currents through them. As such, the geometric designs with small volumes had been eliminated along with sharp edges that experience high heat concentration and static compression and tension forces from the current. The original plan of exposing rods with 2 cm of exposed material was updated, and the lengths of the rods were reduced to 1.5 cm above the DiMES port surface for the final design.

### 3.6 Carbon ablation experiments in DIII-D

DiMES heads #1 and #2 (Table 1 and Figure 2) were inserted in the DIII-D tokamak during the Frontiers in Plasma Science experiment 2019-08-04 "Hypervelocity impact in stellar media: heat shielding, shock fronts and ablation clouds." DiMES head #1 consisted of three carbon rods of different geometries. DiMES head #2 hosted a set of three cylindrical rods, two of which were coated with 30 microns of SiC. Both systems were exposed to the scrape-off layer heat fluxes, with rod #1 in each system located at a smaller radius, closer to the outer strike point of a lower single null, L-mode discharge. Rod #3 in each configuration was located at a larger radius, further away from the outer strike point, and therefore experienced smaller heat fluxes.

Both sets of rods were exposed to high heat fluxes of different pulse lengths during these experiments. DiMES head #1 was exposed for a total time of 6 seconds, while head #2 was exposed for a total of 9 seconds. All the rods in the experiment worked reliably as predicted by our analysis, sustained the applied heat fluxes, exhibited partial ablation as predicted by the models, did not break and did not cause plasma discharge termination by a disruption. Figure 5 shows the visible light imaging of the DiMES head #2 in the DIII-D discharge 186097. Brightness amplitude in the image corresponds to high levels of molecular deuterium emission near the sample surface and is approximately proportional to the surface temperature of the rods. The analysis of the experimental measurements is ongoing and will be presented at the conference and in subsequent publications in the archival literature.

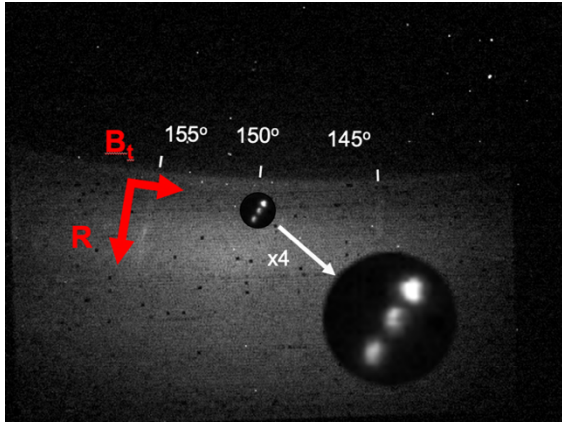

**Figure 5.** Visible light imaging of the DiMES head #2 during exposure to heat fluxes in DIII-D discharge 186097. The direction of the tokamak toroidal field, radial direction, and toroidal angle location in DIII-D are shown for reference. Insert – a zoomed-in (4x) view of the three ablation rods.

## 4. CONCLUSION

Tokamaks, and specifically DIII-D, are able to reproduce environments experienced by probes entering the atmospheres of gas giants within our solar system. In addition to reproducing these environments, the information available due to the advanced diagnostics found in the DIII-D divertor region offer an unrivaled opportunity to not only validate but also improve existing ablation models. Accurate ablation models are essential to the continued development of space exploration. In this work, four types of rods were designed, analyzed, and produced. Limited to the material that can be introduced into a tokamak plasma, ATJ graphite and silicon carbide were chosen as the material for the rods. Three of these rods were made of solid ATJ graphite, while the other had a core of ATJ coated with silicon carbide. Different geometries were produced to determine their impact on the ablation rate. Based on the thermal analysis, the rods are expected to ablate during the experiment, and the amount of material released will be below the upper limit set by DIII-D Operations. In addition, the current-induced forces caused by the local collapse of the Debye sheath due to ablation are not expected to break these rods. This prediction was confirmed during the experimental campaign.

Materials ablation is a multifaceted phenomenon that requires extensive analysis across different areas of engineering and physics. The results of this research and the conducted experiments indicate that tokamak setups are suitable ground testing facilities to replicate material ablation scenarios occurring during high-speed atmospheric entries, which are required in space exploration missions. On a broader level, the current research points to attractive routes for the characterization and study of materials behavior under extreme conditions. The insights developed from the current work and similar studies pave the road for the promising application of ablative materials in rapid development and testing of thermal protection and heat shielding systems.

## ACKNOWLEDGEMENTS

This material is based upon work supported by the U.S. Department of Energy, Office of Science, Office of Fusion Energy Sciences, using the DIII-D National Fusion Facility, a DOE Office of Science user facility, under DE-SC0021338, DE-SC0021620, and DE-FC02-04ER54698. The project was carried out as part of the UCSD MAE156b Fundamental Principles of Mechanical Design II course. The authors would like to thank the DIII-D team for the support of the project and the experiments. The authors would also like to thank R.D. Smirnov for his help with UEDGE simulations.